\documentclass[12pt]{article}

\usepackage[utf8]{inputenc}
\usepackage[T1]{fontenc}

\usepackage{setspace} % must come before hyperref
\usepackage[vmargin = 1.25in, hmargin =1.25in]{geometry}

\usepackage[usenames,dvipsnames,svgnames]{xcolor}
\usepackage{colortbl}
\usepackage{amsmath, amssymb, amsfonts, graphicx, tikz,pdflscape, mathtools, amsthm, upgreek, bm,pgfplots,csvsimple,multirow, multicol, booktabs}
\usepackage{verbatim}
\usepackage{natbib}
\usepackage{accents}

\usepackage{needspace}

\usepackage{comment}
\usepackage[inline]{enumitem}
\usepackage{tocvsec2}
\usepackage{threeparttable}
\usetikzlibrary{calc, cd}
\setlist{noitemsep, topsep=0pt}
\usepackage[skip = 10pt]{caption}
\allowdisplaybreaks[1]
\allowdisplaybreaks[1]

\definecolor{Blue}{RGB}{86,180,233}
\definecolor{Orange}{RGB}{230,159,0}
\definecolor{Green}{RGB}{0,158,115}
\definecolor{GmailBlue}{RGB}{42, 93, 176} % for links
\usepackage[
	pagebackref,
	colorlinks=true,
	citecolor= GmailBlue,
	linkcolor=GmailBlue,
	urlcolor = GmailBlue
]{hyperref}

\newcommand{\bibtexorder}[1]{}

\usepackage{pgfplots}
\usepgfplotslibrary{groupplots,colorbrewer}
\pgfplotsset{compat=newest}
\pgfplotsset{cycle list/Set1}
\usepackage{tikz}
\usetikzlibrary{matrix,calc,shapes,arrows.meta,positioning}
\tikzset{
    vertex/.style = {shape=circle,draw, minimum size = 1.8em, inner sep = 0pt},
    edge/.style = {->,> = latex}
}

\usepackage[capitalize,noabbrev]{cleveref}

\newtheoremstyle{break}% name
{}%         Space above, empty = `usual value'
{}%         Space below
{}% Body font
{}%         Indent amount (empty = no indent, \parindent = para indent)
{\bfseries}% Thm head font
{}%        Punctuation after thm head
{\newline}% Space after thm head: \newline = linebreak
{}%         Thm head spec

\newtheorem*{theorem*}{Theorem}
\newtheorem*{cor*}{Corollary}

\crefname{prop}{Proposition}{Propositions}
\crefname{thm}{Theorem}{Theorems}
\crefname{lem}{Lemma}{Lemmas}
\crefname{blem}{Lemma}{Lemmas}

\theoremstyle{definition}

\newtheorem*{rem*}{Remark}
\newtheorem*{claim*}{Claim}

\def\s{\sigma}

\def\w{\omega}

\def\W{\Omega}

\title{Comment on Matsushima, Miyazaki, and Yagi (2010) ``Role of Linking Mechanisms in Multitask Agency with Hidden Information''\thanks{We thank Koichi Miyazaki for helpful comments.}}

\author{%
	Ian Ball%
	\thanks{Department of Economics, Massachusetts Institute of Technology, ianball@mit.edu.}
	\and 
	Deniz Kattwinkel%
	\thanks{Department of Economics, University College London, d.kattwinkel@ucl.ac.uk.}
}
\date{7 February 2023}

\begin{document}

\maketitle

 We correct a gap in the proof of Theorem 2 in \citet[p.~2248]{MatsushimaEtal2010}. Given $K$ tasks and a finite set $\W$ of private signals for each task, a quota is a map $B \colon \W \to \{1, \ldots, K \}$ satisfying $\sum_{\w \in \W} B(\w) = K$. In the associated quota mechanism, the message space $M$ consists of all $K$-vectors $\hat{\w} = (\hat{\w}_1, \ldots, \hat{\w}_K)$ in $\W^K$ satisfying $\# \{ k : \hat{\w}_k = \w \} = B(\w)$ for each $\w$ in $\W$. Given such a mechanism and a fixed vector $\hat{\w} = (\hat{\w}_1, \ldots, \hat{\w}_K)$ in $\W^K$, a strategy $\s \colon \W^K \to M$ is \emph{cyclic for $\hat{\w}$} if there exists a subset $S$ of $\{1, \ldots, K\}$ with $\# S \geq 2$ and a one-to-one function $\tau \colon \{1, \ldots, \# S\} \to S$ such that (i) $\hat{\w}_s \neq \hat{\w}_{s'}$ for all distinct $s,s' \in S$, and (ii) $\s_{\tau(\ell)} (\hat{\w}) = \hat{\w}_{\tau(\ell+1)}$ for all $\ell$ in $\{1, \ldots, \#S \}$, with $\#S + 1$ defined to equal $1$. 

\citet[p.~2249]{MatsushimaEtal2010} claim that if a strategy $\s$ is not cyclic for a vector $\hat{\w} = (\hat{\w}_1, \ldots, \hat{\w}_K)$ in $ \W^K$, then 
\begin{equation} \label{ineq_false}
    \frac{\# \{ k: \s_k ( \hat{\w}) \neq \hat{\w}_k \}}{K} 
    \leq 
    \sum_{\w \in \W} \, \biggl\vert \frac{\# \{ k: \hat{\w}_k = \w \}}{K} - \frac{B(\w)}{K} \biggr\vert.
\end{equation}
This claim is not correct. Here is a counterexample. Let $\W = \{ A,B,C, D\}$, $K = 4$, and $B(\w) = 1$ for all $\w \in \W$. Consider a strategy $\s$ such that $\s ( A, A, B, C) = (A, B, C, D)$. It is straightforward to check that $\s$ is not cyclic at $(A,A,B,C)$. Taking $\hat{\w} = (A, A, B, C)$ in \eqref{ineq_false}, the left side equals $3/4$ and the right side equals $1/2$, contrary to the inequality.  

To complete the proof of Theorem 2, it suffices to show instead that if $\s$ is not cyclic for a vector $\hat{\w} = (\hat{\w}_1, \ldots, \hat{\w}_K)$ in $\W^K$, then 
\begin{equation} \label{ineq_correct}
    \frac{\# \{ k: \s_k ( \hat{\w}) \neq \hat{\w}_k \}}{K} 
    \leq 
   \frac{ |\W| - 1}{2} \sum_{\w \in \W} \, \biggl\vert \frac{\# \{ k: \hat{\w}_k = \w \}}{K} - \frac{B(\w)}{K} \biggr\vert.
\end{equation} 
Inequality \eqref{ineq_correct} is strictly weaker than \eqref{ineq_false} if $|\W| \geq 4$. 

We prove the contrapositive of this modified claim, following \cite{BJK2022}, which corrects a claim in \cite{jackson2007overcoming} similar to that in \cite{MatsushimaEtal2010}. Fix a strategy $\s$ and a vector $\hat{\w} = (\hat{\w}_1, \ldots, \hat{\w}_K)$ in $\W^K$ such that \eqref{ineq_correct} is violated. By Lemma 2 \citep[p.~o6]{BJK2022}, there exists a nonempty subset $T$ of $\{1, \ldots, K\}$ and a bijection $\pi$ on $T$ such that (i) $(K -  \# T)/K$ is at most the right side of \eqref{ineq_correct}, and (ii) $\s_k (\hat{\w}) = \hat{\w}_{\pi (k)}$ for all $k$ in $T$. Since \eqref{ineq_correct} is violated, there exists $j$ in $T$ such that $\hat{\w}_{\pi(j)} = \s_j (\hat{\w}) \neq \hat{\w}_{j}$. Define a directed multigraph with vertex set $\W$ and edge set indexed by $k$ in $T$, where edge $k$ is from $\hat{\w}_k$ to $\hat{\w}_{\pi(k)}$. This multigraph is balanced, so it can be decomposed into edge-disjoint cycles.\footnote{These cycles can be constructed as follows. Start at a node with an outgoing edge. Form a path by arbitrarily selecting outgoing edges until the path contains a cycle. Remove the cycle and repeat. Since the graph remains balanced after each cycle is removed, this process can terminate only when every edge has been removed.}  The cycle containing edge $j$ has length at least $2$. Let $S$ be the set of indices of the edges in this cycle. Define $\tau$ recursively as follows. Let $\tau(1) = j$. For each $\ell$ in $\{1, \ldots, \#S - 1\}$, let $\tau(\ell + 1)$ be the label of the unique edge in $S$ with tail $\hat{\w}_{\pi(\tau(\ell))}$. By construction, $\hat{\w}_{\tau(\ell + 1)} = \hat{\w}_{\pi ( \tau(\ell))} = \s_{\tau(\ell)} (\hat{\w})$ for each $\ell$ in $\{1, \ldots, \# S\}$, with $\#S + 1$ defined to equal $1$. Therefore, $\s$ is cyclic at $\hat{\w}$. 

%In particular, this implies that the right side of \eqref{ineq_correct} is strictly less than $1$. 
\begin{comment}
\footnote{If for each $\hat{\w}$ in $\W^K$, strategy $\s$ is not cyclic at $\hat{\w}$, then $\s$ is \emph{permutation-truthful}, as defined in \cite{BJK2022}. It follows from  that $\s$ satisfies \eqref{ineq_correct} for each $\hat{\w}$ in $\W^K$.}
\end{comment}
\bibliographystyle{ecta}
\bibliography{lit.bib}

\end{document}